\documentclass[aps,pra,reprint,amsmath,amssymb,longbibliography]{revtex4-1}
\usepackage{graphicx}
\usepackage{bm}
\usepackage[colorlinks=true,linkcolor=blue,citecolor=blue,urlcolor=blue]{hyperref}
\hypersetup{pdftitle={Two-qubit binegativity can increase under local amplitude damping}}
\newcommand{\Tr}{\operatorname{Tr}}
\newcommand{\Ntwo}{N_{2}}
\newcommand{\ket}[1]{\lvert #1\rangle}
\newcommand{\bra}[1]{\langle #1\rvert}
\newcommand{\proj}[1]{\ket{#1}\bra{#1}}
\newcommand{\diag}{\operatorname{diag}}

\begin{document}
\title{Two-qubit binegativity can increase under local amplitude damping}
\author{Xiao-Lan Zong}
\author{Wei Song}
\email{wsong1@mail.ustc.edu.cn}
\affiliation{School of Physics and Materials Engineering, Hefei Normal University, Hefei 230601, China}
\author{Ming Yang}
\affiliation{School of Physics, Anhui University, Hefei, 230601, China}
\affiliation{Leibniz International Joint Research Center of Materials Sciences of Anhui Province, Anhui University, Hefei, 230601, China}

\begin{abstract}
It has been conjectured that the binegativity of two-qubit states is an entanglement monotone.
We present a full-rank state as a counterexample: applying amplitude damping to only one of the
qubits can increase the binegativity of the state. The output state remains full rank, and the amount
of increase can be calculated exactly. In addition, we construct a family of counterexamples that
vary continuously with a parameter. In the corresponding one-way local operations and classical communication protocol, the
probability-weighted average of the binegativity over the branches after the local measurement is also
higher than that of the initial state.
\end{abstract}
\maketitle

\section{Introduction}
An entanglement monotone must not increase on average under local operations and
classical communication (LOCC)~\cite{Vidal2000}. For two-qubit states, both
concurrence and negativity have explicit expressions, but the orderings of entanglement strength obtained
according to these two measures are not always consistent~\cite{Wootters1998,VidalWerner2002,Verstraete2001}.
Girard and Gour proposed binegativity and conjectured that it is monotonic under both LOCC and
positive-partial-transpose-preserving (PPT) channels~\cite{GirardGour2017}.

An earlier proof of this conjecture was withdrawn in the third version of~\cite{GirardGour2017},
and the associated construction of entanglement conversion witnesses was also revised~\cite{GirardGour2015}.
However, these revisions did not provide a counterexample. Subsequently, studies of a specific class of quantum
states under amplitude damping, phase damping, and depolarizing noise found that the behavior of binegativity
was consistent with this conjecture~\cite{SazimAwasthi2018}. But this does not
show that the conjecture holds for arbitrary two-qubit states.

This paper provides a counterexample that disproves this conjecture. We construct a full-rank two-qubit state
such that applying amplitude damping to only one of the qubits increases the binegativity of the state, while
the output state remains full rank. This increase can be verified directly by calculating the negative eigenvalues
of the partial transpose and their corresponding eigenvectors before and after the evolution. This analysis
also explains why the initial states chosen in previous noise studies cannot produce counterexamples in which
the binegativity exceeds its initial value.

We also construct a family of counterexamples that vary continuously with a parameter by using a one-way
LOCC protocol. In this protocol, even if only local measurements are performed, the average binegativity over
the branches after measurement is higher than that of the initial state. The appendix gives a full-rank counterexample
under the same protocol.

\section{Preliminaries}
Throughout, $\Gamma$ denotes partial transposition with respect to Bob's qubit. For a Hermitian operator $X$, its
negative part is denoted by $X_-=(|X|-X)/2$; under this definition, $X_-$ is a positive semidefinite operator.
The normalization convention for negativity adopted in this paper is
$N(\rho)=\|\rho^\Gamma\|_1-1=2\Tr[(\rho^\Gamma)_-]$. The binegativity is defined as~\cite{GirardGour2017}
\begin{equation}
 \Ntwo(\rho)=\Tr[(\rho^\Gamma)_-]
 +2\Tr\!\left[\left(((\rho^\Gamma)_-)^\Gamma\right)_-\right].
 \label{eq:def}
\end{equation}
This definition also applies to unnormalized positive semidefinite operators and satisfies positive homogeneity.
Equation~\eqref{eq:def} defines a scalar and should be distinguished from the operator $|\rho^\Gamma|^\Gamma$.
Ishizaka has shown that the latter is positive semidefinite for any two-qubit state~\cite{Ishizaka2004}.

The partial transpose of an entangled two-qubit state has exactly one negative eigenvalue~\cite{Verstraete2001}.
Denote it by $-t$, and denote the corresponding normalized eigenvector by $\ket v$.
Then $(\rho^\Gamma)_-=t\proj v$. Substituting into Eq.~\eqref{eq:def}, we
obtain the known formula~\cite{GirardGour2017,SazimAwasthi2018}
\begin{equation}
 \Ntwo(\rho)=t[1+C(v)]=\frac{N(\rho)}2[1+C(v)],
 \label{eq:spectral}
\end{equation}
where $C(v)$ is the concurrence of the pure state $\ket v$. In particular, we have $\Ntwo(\rho)\leq N(\rho)$.

The following example will use a positive semidefinite operator having the following form in the basis
$00,01,10,11$:
\begin{equation}
 R=\begin{pmatrix}
 a&0&0&w\\0&b&0&0\\0&0&c&0\\w&0&0&d
 \end{pmatrix},\qquad R\geq0,\quad w>0.
 \label{eq:X}
\end{equation}
Here $R$ is not required to have trace $1$. When $w^2>bc$,
the negative eigenvalue of $R^\Gamma$ comes from the subblock
$\left(\begin{smallmatrix}b&w\\w&c\end{smallmatrix}\right)$.
Let $s=\sqrt{(b-c)^2+4w^2}$. Then the absolute value of this negative eigenvalue is $(s-b-c)/2$, and the
concurrence of the corresponding normalized eigenvector is $2w/s$. Therefore,
\begin{equation}
 \Ntwo(R)=\frac{s-b-c}{2}\left(1+\frac{2w}{s}\right).
 \label{eq:block}
\end{equation}
When $w^2\leq bc$, we have $\Ntwo(R)=0$.

\section{A counterexample under local amplitude damping}
Consider the following two-qubit state:
\begin{equation}
 \rho_* =\frac1{8700}
 \begin{pmatrix}
 8000&0&0&88\\0&3&0&0\\0&0&696&0\\88&0&0&1
 \end{pmatrix}.
 \label{eq:input}
\end{equation}
The trace of this matrix is $1$. The two one-dimensional blocks are positive; the diagonal entries of the two-dimensional
block are both positive, and $8000-88^2=256>0$, so it is also positive definite. Therefore, $\rho_*$ is a
full-rank quantum state.

Let Alice's qubit pass through an amplitude damping channel with damping probability $\eta_*=253/289$.
The Kraus operators of this channel are
\begin{equation}
 A_0=\diag(1,6/17),\qquad
 A_1=\frac{\sqrt{253}}{17}\ket0\bra1,
 \label{eq:AD}
\end{equation}
which satisfy $\sum_jA_j^\dagger A_j=I$. The output state after the action is
\begin{align}
 \sigma_*&=\sum_j(A_j\otimes I)\rho_*(A_j^\dagger\otimes I)\notag\\
 &=\frac1{628575}
 \begin{pmatrix}
 622022&0&0&2244\\0&280&0&0\\0&0&6264&0\\2244&0&0&9
 \end{pmatrix}.
 \label{eq:output}
\end{align}
The trace of the output state remains $1$. All its diagonal entries are positive, and
$622022\cdot9-2244^2=562662>0$, so $\sigma_*$ is also positive definite and remains a full-rank state.

Direct calculation of the partial transposes of the input and output states gives their respective negative eigenvalues
and corresponding normalized eigenvectors:
\begin{align}
 \lambda_-(\rho_*^\Gamma)&=-\frac2{2175},&
 \ket{v_{\rho_*}}&=\frac{8\ket{01}-\ket{10}}{\sqrt{65}},
 \label{eq:inputpair}\\
 \lambda_-(\sigma_*^\Gamma)&=-\frac{156}{209525},&
 \ket{v_{\sigma_*}}&=\frac{3\ket{01}-\ket{10}}{\sqrt{10}}.
 \label{eq:outputpair}
\end{align}
The concurrences of the two eigenvectors are $16/65$ and $3/5$, respectively.
Substituting into Eq.~\eqref{eq:spectral}, we obtain
\begin{align}
 \Ntwo(\rho_*)&=\frac{54}{47125},&
 \Ntwo(\sigma_*)&=\frac{1248}{1047625},\notag\\
 \Ntwo(\sigma_*)-\Ntwo(\rho_*)&=\frac{618}{13619125}>0.
 \label{eq:gain}
\end{align}
It can be seen that the binegativity of the output state is $2704/2601$ times that of the initial state, an increase of
about $3.96\%$ over the initial value. This process involves only a deterministic local operation on a single qubit
and requires no postselection, and therefore constitutes a counterexample showing that binegativity does not satisfy
monotonicity under LOCC. Since LOCC channels are PPT channels, this counterexample also disproves the
monotonicity of binegativity under PPT channels.

Furthermore, by continuity, as long as the initial state is sufficiently close to $\rho_*$, the same channel will still
strictly increase its binegativity, and both the input and output states remain full rank.

\begin{figure}[t]
 \centering
 \includegraphics[width=\columnwidth]{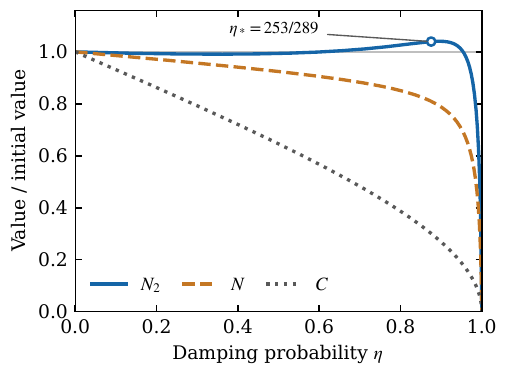}
 \caption{Variation of the binegativity, negativity, and concurrence with the damping probability when amplitude damping
 is applied only to Alice's qubit, starting from the state in Eq.~\eqref{eq:input} as the initial state.
 Each quantity is normalized by its respective initial value. The circle corresponds to $\eta_*=253/289$,
 at which the binegativity is already higher than its initial value.}
 \label{fig:damping}
\end{figure}

Why the binegativity increases can be seen from Eq.~\eqref{eq:spectral}.
The absolute value of the negative eigenvalue decreases to $6786/8381$ of its original value, while the
factor $1+C(v)$ increases to $104/81$ of its original value; the product of these two multiplicative factors is greater than $1$.
That is, although the negativity decreases, the entanglement of the corresponding eigenvector increases,
ultimately causing the binegativity to increase. Figure~\ref{fig:damping} shows the variation of the binegativity,
negativity, and concurrence with the damping probability; it can be seen that the concurrence also decreases.

Next, we compare this counterexample with the initial states used in Ref.~\cite{SazimAwasthi2018}.
The initial states in that paper have the following form: $\rho_{\mathrm{EW}}=(1-p)I_4/4+p\proj\Psi$,
where $\ket\Psi=\alpha\ket{00}+\beta\ket{11}$ and $|\alpha|^2+|\beta|^2=1$.
For this family of states, once the partial transpose has a negative eigenvalue, the corresponding eigenvector
is a maximally entangled state, so $\Ntwo(\rho_{\mathrm{EW}})=N(\rho_{\mathrm{EW}})$.
When the state is separable, both the binegativity and the negativity are zero, and the equality also holds.

For any input state satisfying $\Ntwo(\rho)=N(\rho)$ and any PPT channel $\Lambda$ whose output is a two-qubit state,
from the monotonicity of negativity~\cite{VidalWerner2002,Plenio2005} we obtain
\begin{equation}
 \Ntwo(\Lambda(\rho))\leq N(\Lambda(\rho))
 \leq N(\rho)=\Ntwo(\rho).
 \label{eq:protected}
\end{equation}
Therefore, starting from such initial states, the binegativity cannot exceed its initial value after the action of
the above channels, and thus no such counterexample can be obtained. However, this argument only compares
the initial state with the output state and does not prove monotonicity between any two intermediate times during
the evolution. For the initial state in Eq.~\eqref{eq:input}, the binegativity and the negativity are not equal;
instead, they satisfy $\Ntwo(\rho_*)/N(\rho_*)=81/130<1$.

\section{A family of counterexamples}
Next, we construct a family of counterexamples that vary continuously with a parameter via a one-way LOCC
protocol. Let $D_z=\diag(1,z)$, where $0<z<1$, and define
\begin{align}
 \mathcal L_z(R)={}&(D_z\otimes I)R(D_z\otimes I)\notag\\
 &+\Tr\!\left[((I-D_z^2)\otimes I)R\right]\proj{00}.
 \label{eq:reset}
\end{align}
This channel can be implemented as follows. Alice first performs a local measurement, with measurement operators
$M_0=D_z$ and $M_1=\sqrt{1-z^2}\proj1$.
If the outcome is $0$, both parties keep the post-measurement system; if the outcome is $1$, Alice informs Bob of the result,
and both parties reset their qubits to $\ket0$. Ignoring the measurement outcome then yields the trace-preserving channel
in Eq.~\eqref{eq:reset}.

Consider the family of states studied in Refs.~\cite{Verstraete2001,GirardGour2017}:
\begin{equation}
 \rho_p=p\proj{\Phi^+}+(1-p)\proj{10},\qquad 0<p<1,
 \label{eq:family}
\end{equation}
where $\ket{\Phi^+}=(\ket{00}+\ket{11})/\sqrt2$. Let $r=(1-p)/p$.
From Eq.~\eqref{eq:block}, the binegativity of the input and output states is given by
\begin{align}
 \Ntwo(\rho_p)&=\frac{p}{2r}h(r),\notag\\
 \Ntwo(\mathcal L_z(\rho_p))&=\frac{p}{2r}h(rz),
 \label{eq:familyvalues}
\end{align}
where
\begin{equation}
 h(u)=u\bigl(\sqrt{1+u^2}-u\bigr)
 \left(1+\frac1{\sqrt{1+u^2}}\right).
 \label{eq:h}
\end{equation}
To explain the second equality, it suffices to note that the branch corresponding to measurement outcome $0$ satisfies
$b=0$, $c=(1-p)z^2$, and $w=pz/2$. The reset term only changes the matrix element $a$, and therefore does not
affect the result of the calculation in Eq.~\eqref{eq:block}.

Take
\begin{equation}
 0<p<\frac37,\qquad z=\frac4{3r}.
 \label{eq:choice}
\end{equation}
Then $r>4/3$ and $0<z<1$. To prove that the binegativity increases, it suffices to compare $h(r)$ with $h(4/3)$.
Let $v=\sqrt{1+u^2}-u$. For $u\geq4/3$, we have $0<v\leq1/3$, and
\begin{align}
 h(u)&=g(v)=\frac{(1-v)(1+v)^3}{2(1+v^2)},\notag\\
 g'(v)&=\frac{(1+v)^2(1-3v+v^2-v^3)}{(1+v^2)^2}>0.
 \label{eq:derivative}
\end{align}
Since $1-3v\geq0$ and $v^2-v^3>0$, the derivative in the above expression is strictly positive.
Moreover, because $v$ decreases as $u$ increases, $h$ is strictly decreasing on $[4/3,\infty)$.
Combining this with $h(4/3)=32/45$, we obtain
\begin{equation}
 \Ntwo(\mathcal L_z(\rho_p))-\Ntwo(\rho_p)
 =\frac{p}{2r}\left[\frac{32}{45}-h(r)\right]>0.
 \label{eq:familygain}
\end{equation}
For example, taking $p=5/17$ and $z=5/9$ gives
\begin{equation}
 \Ntwo(\rho_p)=\frac9{221},\qquad
 \Ntwo(\mathcal L_z(\rho_p))=\frac{20}{459},
 \label{eq:rational}
\end{equation}
and the binegativity increases by $1/351$.

The above calculation also shows that, even without the reset, Alice's local measurement alone can make
the probability-weighted average of the binegativity over the branches higher than its initial value.
Let $R_j=(M_j\otimes I)\rho_p(M_j^\dagger\otimes I)$, $q_j=\Tr R_j$, and $\rho_j=R_j/q_j$.
For measurement outcome $1$,
\begin{equation}
 R_1=(1-z^2)\proj1\otimes
 \left[(1-p)\proj0+\frac p2\proj1\right],
 \label{eq:failure}
\end{equation}
which is a separable branch with zero binegativity. Meanwhile, the matrix elements $b,c,w$ of $R_0$ are the same as
the corresponding matrix elements of $\mathcal L_z(\rho_p)$. Therefore, by positive homogeneity and Eq.~\eqref{eq:block}, we obtain
\begin{equation}
 \sum_jq_j\Ntwo(\rho_j)=\Ntwo(R_0)
 =\Ntwo(\mathcal L_z(\rho_p))>\Ntwo(\rho_p).
 \label{eq:average}
\end{equation}
This inequality holds as long as the parameters satisfy Eq.~\eqref{eq:choice}, which shows that binegativity also fails to satisfy
monotonicity in the averaged sense. Appendix~\ref{app:fullrank} gives a full-rank example under the same protocol, in
which the binegativity exhibits the same two kinds of non-monotonicity.

\section{Concluding remarks}
We prove that the binegativity of two-qubit states does not satisfy monotonicity under local operations.
We first give a full-rank state as a counterexample: applying amplitude damping to only one of the qubits can increase
the binegativity. Then, via a one-way LOCC protocol, we construct a family of counterexamples that vary continuously
with a parameter, and show that local measurements can also increase the binegativity in the averaged sense.

\appendix
\section{A full-rank example for the one-way protocol}
\label{app:fullrank}
The protocol in Eq.~\eqref{eq:reset} can also increase the binegativity of a full-rank state.
Take $z=4/7$ and choose the input state
\begin{equation}
 \widehat\rho=\frac1{4420}
 \begin{pmatrix}
 990&0&0&495\\0&32&0&0\\0&0&2408&0\\495&0&0&990
 \end{pmatrix}.
 \label{eq:appendixinput}
\end{equation}
The four eigenvalues of this state are $495/4420$, $1485/4420$, $32/4420$, and $2408/4420$, all of which are positive,
so $\widehat\rho>0$. Since $D_{4/7}$ is invertible, the branch corresponding to measurement outcome $0$ remains positive
definite, and hence the output state $\widehat\sigma=\mathcal L_{4/7}(\widehat\rho)$ is also positive definite.
From Eq.~\eqref{eq:block}, we obtain
\begin{align}
 \Ntwo(\widehat\rho)&=\frac{603}{28730},&
 \Ntwo(\widehat\sigma)&=\frac{872}{38675},\notag\\
 \Ntwo(\widehat\sigma)-\Ntwo(\widehat\rho)
 &=\frac{1567}{1005550}>0.
 \label{eq:appendixgain}
\end{align}
On the other hand, since $M_1$ has rank $1$, the branch corresponding to measurement outcome $1$ is separable,
and its binegativity is zero. Adding the reset term to the branch corresponding to outcome $0$ only changes the
matrix element $a$ in Eq.~\eqref{eq:X} and does not change the binegativity.
Therefore, following the same derivation as in Eq.~\eqref{eq:average}, by positive homogeneity and Eq.~\eqref{eq:block}, the
probability-weighted average of the binegativity over the branches after Alice's measurement is exactly equal to
$\Ntwo(\widehat\sigma)$. This shows that, for this full-rank input state, local measurement also increases the binegativity
in the averaged sense.

\acknowledgments
This work was supported by the Natural Science Foundation of Anhui Province under Grant No. 2508085ZD001, and Scientific Research Projects of Anhui Provincial Department of Education under Grant Nos. 2025AHGXZK20126 and 2025AHGXZK50054.

\end{document}